\documentclass[pra,aps,twocolumn,tightenlines,10pt,balancelastpage]{revtex4-1}%
\usepackage[utf8x]{inputenc}

\usepackage{graphics}
\usepackage{epsfig}
\usepackage{subfigure}
\usepackage{rotating}

\usepackage{amsfonts}
\usepackage{amsmath}
\usepackage{amssymb}
\usepackage{placeins}
\usepackage{xcolor}
\usepackage{soul}
\usepackage{fixmath}
\usepackage[colorlinks]{hyperref}

\begin{document}

\title{Temporal cavities as temporal mode filters for frequency combs} %
\author{B. Dioum$^1$, S. Srivastava$^1$, M. Karpi\'nski$^2$ and G. Patera$^1$}
\address{
$^1$Univ. Lille, CNRS, UMR 8523 -- PhLAM -- Physique des Lasers Atomes et Mol\'{e}cules, F-59000 Lille, France
\\
$^2$ Faculty of Physics, University of Warsaw, Pasteura 5, 02-093 Warszawa, Poland
}

\begin{abstract}
Broadband temporal modes of pulsed optical fields have been recently recognized as very promising for photonic quantum information processing and time-frequency metrology. 
Exploiting their full potential demands efficient and flexible tools for their manipulation. Among the tools demonstrated surprisingly the most basic, a single-mode temporal filter, is missing. 
In this work, we propose an experimentally feasible approach to realize a genuine single-mode temporal filter that is based on the concept of temporal cavity, 
a device with temporal mode-dependent resonances in the basis of frequency comb modes. This functionality is achieved as temporal-domain analogy of spatial-mode cleaner cavities.  
This device will enable robust  temporal mode filtering and detection, opening new prospects in time-frequency metrology and multidimensional quantum information processing.
\end{abstract}

\date{\today}
\maketitle

Temporal modes (TMs) of optical ultra-short pulses provide a reliable and flexible encoding for photonic non-classical states~\cite{Brecht2015}. 
Their noiseless manipulation plays an important role in the processing of quantum information and in communication networks \cite{Raymer2012,Humphreys2014,Raymer2020,Karpinski2021},
it is therefore important to have schemes for sorting and distributing them within a network. 
This would enable the implementation of protocols for multipartite entanglement sharing, quantum teleportation, quantum key distribution \cite{Brecht2015,Raymer2012,Wang2022}, 
as well as for the synthesis of multimode entangled states such as cluster states~\cite{Patera2012,Yokoyama2013,Cai2017}, 
a fundamental resource for the implementation of a computation model based on projective measurements~\cite{Menicucci2007}.

Mode-selective techniques have been developed in the last few years such as, for example, the quantum pulse gate (QPG)~\cite{Brecht2011, Eckstein2011, Reddy2015, Reddy2018}. 
Whereas these devices have the capability of sorting the modal content of the input while preserving its quantum properties \cite{Serino2022}, 
they rely on the use of a reference pump signal in the desired TM and an engineered phase-matching that inherently modify both the central wavelength and the selected TM mode 
and that may not be availbale for arbitrary carrier frequencies.
On the other side, it is not known whether the complex approach of multi-stage arbitrary modulations of spectral and temporal phases~\cite{Ashby2020,Joshi2022} could provide genuine filtering capability. 
To date, an approach to implement a genuine filter, which would sort TMs while maintaining their carrier frequency as well as their original shape (in time or frequency domain), has not been shown. 

In this work, we propose the ``temporal cavity" as a device that is able to perform genuine TM filtering. 
We concieved this scheme by using the principles of \textit{space-time duality}~\cite{Kolner1994}, i.e.
the formal analogy between the spatial and temporal degrees of freedom of propagating optical beams. 
One of the most fascinating results yield by this analogy is the time lens, a device that mimics, in the time domain, the effect of thin lenses on spatial images. 
Time lenses have been implemented with deterministic linear electro-optical phase modulation~\cite{Nakazawa1998,Azana2004,Karpinski2016,Sosnicki2020,Karpinski2021} or 
non-linear optical processes~\cite{Bennett1994,Bennett1999,Foster2008,Foster2009,Okawachi2009,Kuzucu2009} and 
have been adopted for manipulation of pulses in discrete~\cite{Zhu2013,Donohue2015,Karpinski2016,Mittal2017,Sosnicki2018,Shi2020} or 
in continuous~\cite{Patera2015,Patera2017,Shi2017,Patera2018,Shi2018} variables.

A temporal cavity is obtained by using the time-frequency analog of a spatial-mode cleaning cavity,
a cavity whose eigenvectors are Hermite-Gaussian transverse modes, with mode-dependent reflection and transmission coefficients~\cite{foundphot}.
This effect is due to a mode-dependent spatial Gouy phase accumulated in a round trip.
Spatial-mode cleaners have been employed for reducing the noise induced by the fluctuations of the laser beam profile in gravitational interferometers~\cite{Rudiger1981}, 
for synthesizing spatial multimode quantum beams~\cite{Treps2003, Armstrong2012} and, more recently, they have been proposed as a spatial-mode sorters~\cite{dosSantos2021}, 
which may enable super-resolved imaging \cite{Tsang2018}.

We show that temporal cavities possess resonances generated by the combination of a cavity build-up effect and the temporal Gouy phase shift~\cite{Nakazawa1998}
and that are distinct with respect to the family of Hermite-Gaussian TMs (HGTM). 
Fundamentally different from other TM sorting approaches, temporal cavities not only behave as genuine filters but are also device-independent, 
in the sense that their operation is based only on the principles of temporal imaging and it does not depend 
of whether they are implemented by nonlinear processes or electro-optic phase modulators \cite{Torres-Company2011}.
%

\begin{figure}[t!]
\centering
\includegraphics[width=0.8\columnwidth]{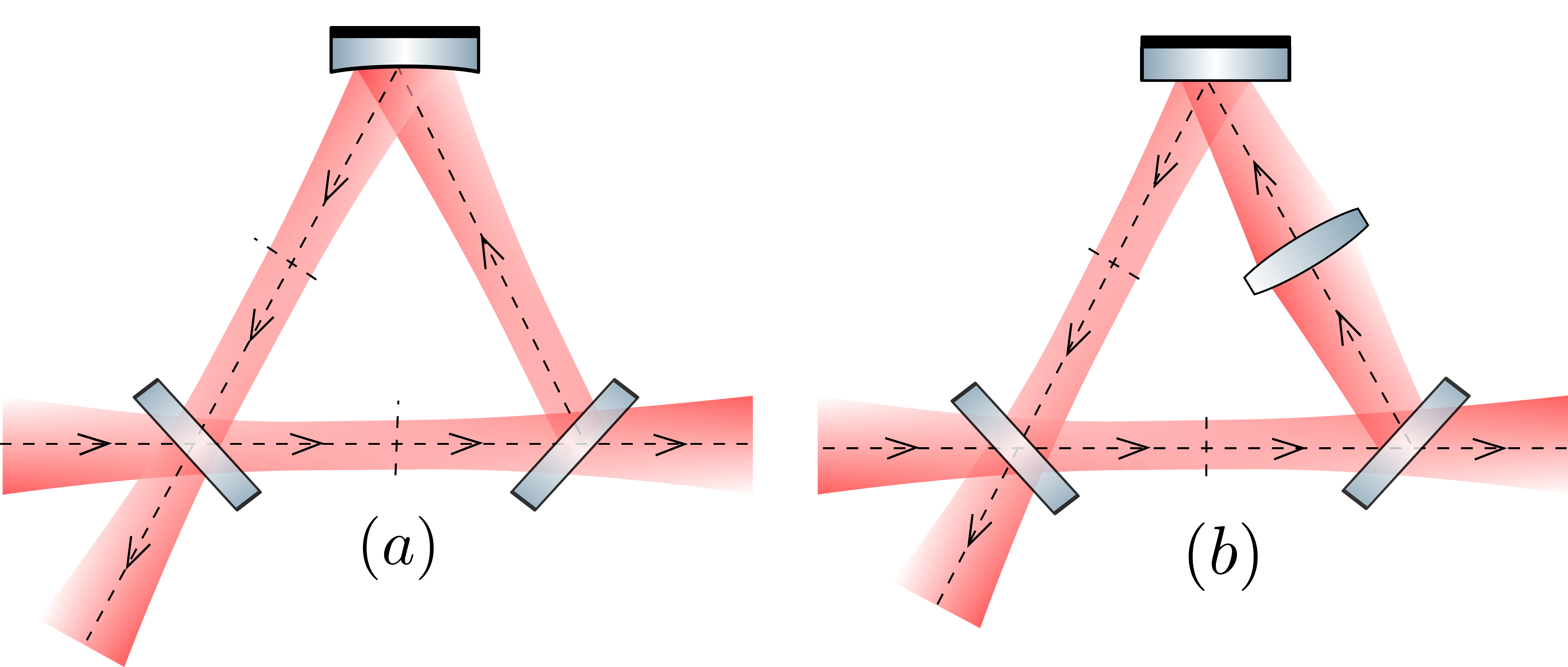}
\caption{(a) Spatial mode-cleaner with spherical mirror, (b) spatial mode-cleaner with a thin lens replacing the spherical mirror.}\label{fig:space cav}
\end{figure}

\indent
\textbf{Temporal cavity.}
When a paraxial monochromatic beam impinges on a spatial-mode cleaner that is tuned on a particular transverse mode, its modal content at resonance with the cavity is transferred, 
while the not resonant modal content is reflected. A typical design for a spatial-mode cleaner cavity consists of two flat partially reflecting mirrors (the input and output couplers) and 
one spherical perfectly reflecting mirror arranged in triangular geometry as in Fig.~\ref{fig:space cav}(a).
%
%
\begin{figure}[t!]
\centering
\includegraphics[width=0.8\columnwidth]{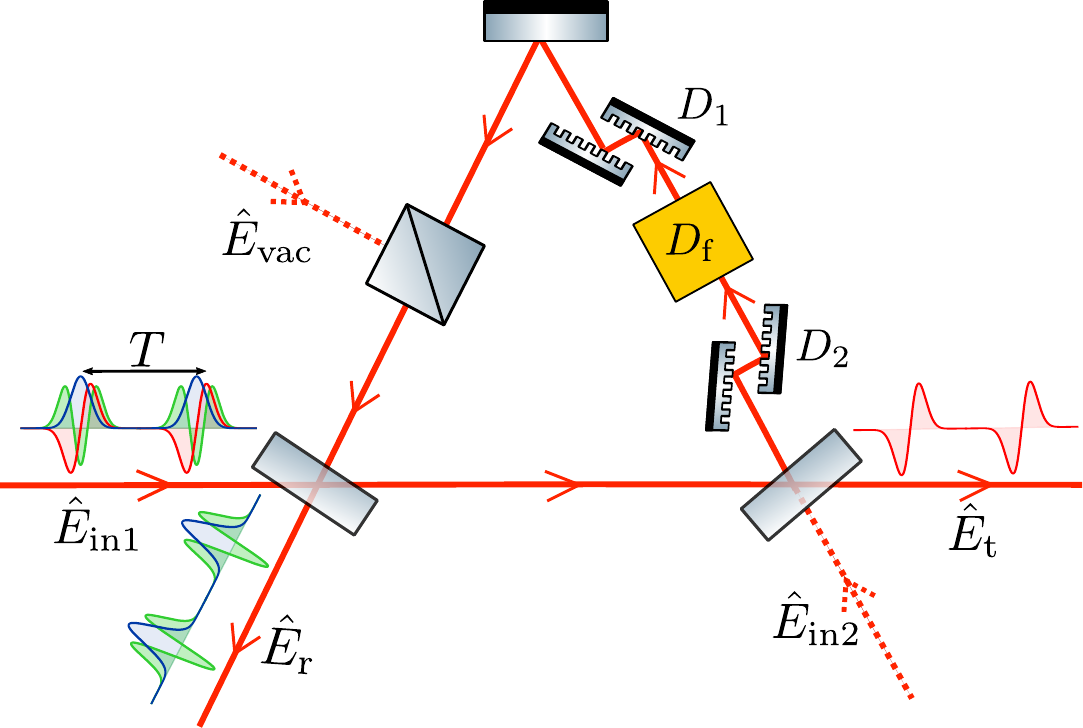}
\caption{Schematic of a temporal cavity with two inputs $\hat{E}_{\mathrm{in1}}$ and $\hat{E}_{\mathrm{in2}}$ and two outputs $\hat{E}_{\mathrm{r}}$ and $\hat{E}_{\mathrm{t}}$. 
The two couplers are characterized by real reflection and transmission coefficients $r_1,r_2$ and $t_1,t_2$. 
A third coupler with reflection coefficient $\rho$ is considered in order to model possible sources of losses during a round-trip. 
The two dispersive elements (two gratings in the picture) with equivalent GDD $D_1$ and $D_2$. A time lens of focal GDD $D_{\mathrm{f}}$ is inserted between the two dispersive elements. 
Note that, contrary to the spatial cavity, the manipulations are done only on the optical pulses. 
The spatial mode of the beam remains the same through propagation in the temporal cavity.}
\label{fig:time cav}
\end{figure}
%

We use the space-time duality~\cite{Kolner1994} in order to translate the triangular cavity to the time domain.
Accordingly, the dispersive propagation of a short pulse in the quasi-monochromatic approximation is the dual, in time domain, of the diffractive evolution of a paraxial monochromatic beam. 
As a consequence, the free space propagation in Fig.~\ref{fig:space cav} must be replaced by the propagation of a pulse in a dispersive medium.
In the quasi-monochromatic approximation, any pulse can be conveniently decomposed on the basis of HGTM. Indeed their evolution and manipulation are
described by the temporal complex pulse parameter $q$, encoding the chirp and the pulse duration, and the temporal ABCD matrix formalism~\cite{Nakazawa1998,Torres-Company2011}. 
See also Appendix~\ref{gauss imag} for more details.

The flat mirrors do not have any special role apart from deflecting the trajectory of the optical beam without changing its shape. 
These elements are kept in the time domain framework with no need for translation. On the contrary, the spherical mirror plays a major role in the emergence of mode-selective resonances. 
Despite ``time reflection" being a possible phenomenon~\cite{Zhang2021}, at the best of our knowledge there is no practical way of implementing a spherical mirror in the time domain. 
This issue can be resolved by observing that its ABCD matrix is formally similar to that of a thin lens whose dual, in time domain, is a time lens which, instead, 
can be easily implemented in experiments \cite{Torres-Company2011,Salem2013,Karpinski2021}. The alternative scheme for a spatial-mode cleaner involves, thus, 
three flat mirrors (two partially reflecting, one perfectly reflecting) and one thin lens as represented in fig.~\ref{fig:space cav}(b). 
The temporal cavity is obtained as the equivalent of this alternative scheme and it is drawn in fig.~\ref{fig:time cav}. 
The beam splitters, with real reflection and transmission coefficients $r_1,t_1$ and $r_2,t_2$, implement the input and output couplers. 
The perfectly reflecting flat mirrors deflect the trajectory in order to create a loop. The dispersive elements are characterized by a total Group Delay Dispersion (GDD) $D_1$ and $D_2$, respectively. 
Notice that, in principle, a scheme with only one dispersive element is also possible, but for generality 
we consider two elements. The time lens is placed between the dispersive elements and it is characterized by a focal-GDD of $D_{\mathrm{f}}$. 
The physical process used for its implementation is not important as far as it does not change the carrier frequency of the pulse. 
Therefore, for this application, the best choice would be electro-optic modulators \cite{Nakazawa1998,Azana2004,Karpinski2016,Sosnicki2020,Karpinski2021} or processes such as cross-phase modulation \cite{Matsuda2016,Hirooka2008a}.
The translation to the time domain is, finally, completed by considering the input beam. 
For mode-cleaners, this is a paraxial monochromatic beam while its dual, in time domain, is a quasi-monochromatic pulse. 
However, in order to get the typical build-up effect of resonant cavities, it is necessary that at the input coupler, 
the $k$-th pulse after one round-trip interferes with a new pulse, say the $(k+1)$-th, entering the cavity. 
As a consequence a temporal cavity can be realized only when impinged by a train of pulses such that their repetition period $T$ matches the round-trip time of the loop (or it is a sub-multiple of it):
\begin{align}
T = \frac{L_{\mathrm{loop}}}{c},
\label{T}
\end{align}
where $L_{\text{loop}}=d_0+d_1+d_2$ is the cavity optical length, $d_0$ is the distance (in empty space) between the input and output couplers and $d_i$ (for $i=1,2$) 
are the optical lengths of the two dispersive media.

A second condition necessary for assuring an interference-based cavity build-up effect is that the duration and chirp of the circulating pulse at loop $k$ matches that of 
the $(k+1)$-th incoming pulse at the input coupler. This is formally expressed by requiring that the $q$-parameter
remains unchanged after each round trip $q_{k+1}=q_k$ (say $q$). By using the temporal ABCD formalism, this condition leads to the eigenvalue condition
\begin{align}
q=\frac{A_{\text{loop}} q + B_{\text{loop}}}{C_{\text{loop}} q + D_{\text{loop}}}
\label{eval eq}
\end{align}
where $A_{\text{loop}}=1-D_2/D_{\mathrm{f}}$, $B_{\text{loop}}=D_{\mathrm{tot}}-D_1 D_2/D_{\mathrm{f}}$, $C_{\text{loop}}=-1/D_{\mathrm{f}}$ and 
$D_{\text{loop}}=1-D_1/D_{\mathrm{f}}$ are the elements of the ABCD matrix of the pulse propagation after one round-trip and $D_{\mathrm{tot}}=D_1+D_2$ is the total GDD. 
The solution of~\eqref{eval eq}
\begin{align}
q=\frac{D_2-D_1}{2}\pm\mathrm{i}\sqrt{D_{\mathrm{tot}}\left(D_{\mathrm{f}}-\frac{D_{\mathrm{tot}}}{4}\right)}
\label{evals}
\end{align}
defines the temporal $q$-parameter that a train of pulses must have in order to be a cavity eigenvector or, conversely, 
it defines the parameters that the temporal cavity needs to have in order to sustain the train of pulses with a given $q$-parameter. 
In particular the stability of the resonator is obtained for $|D_{\mathrm{tot}}|\leq 4 |D_{\mathrm{f}}|$. 
From eq.~\eqref{evals}, it is clear that with two dispersion elements it is possible to arrange the setup so that $\mathrm{Re}[q]=0$ when $D_1=D_2$. 
This allows considering the simpler situation of not chirped input pulses for practical purposes.

Equations~\eqref{T} and~\eqref{evals} represent the two fundamental equations for the temporal cavity. When they are both satisfied, the input and output fields can thus be written as
\begin{align}
\hat{E}^{(+)}(t,z)&=\mathcal{E}_0
\mathrm{e}^{\mathrm{i}\left(k_0 z-\omega_0 t\right)}
\hat{A}_{\mathrm{f}}(t,z),
\label{E+}
\\
\hat{A}_{\mathrm{f}}(t,z)&=
\sum_{k=-\infty}^{+\infty}\sum_{m\in\mathbb{N}} \hat{a}_{\mathrm{f},m}^{(k)} u_m(t-kT,z),
\label{A train}
\end{align}
where the indexes $\mathrm{f}=\{\mathrm{in}1,\mathrm{r},\mathrm{in}2,\mathrm{t}\}$ identify, respectively, the input and the reflected beams at the first coupler, 
the input and the transmitted beams at the second coupler (see fig.\ref{fig:time cav}). 
The modal functions $u_m(t-kT,z)$ form a complete orthonormal set of HGTM that are non-zero in the interval $(k-1)T<t\leq kT$.
\begin{figure}[t!]
    \centering
    \includegraphics[width=0.8\columnwidth]{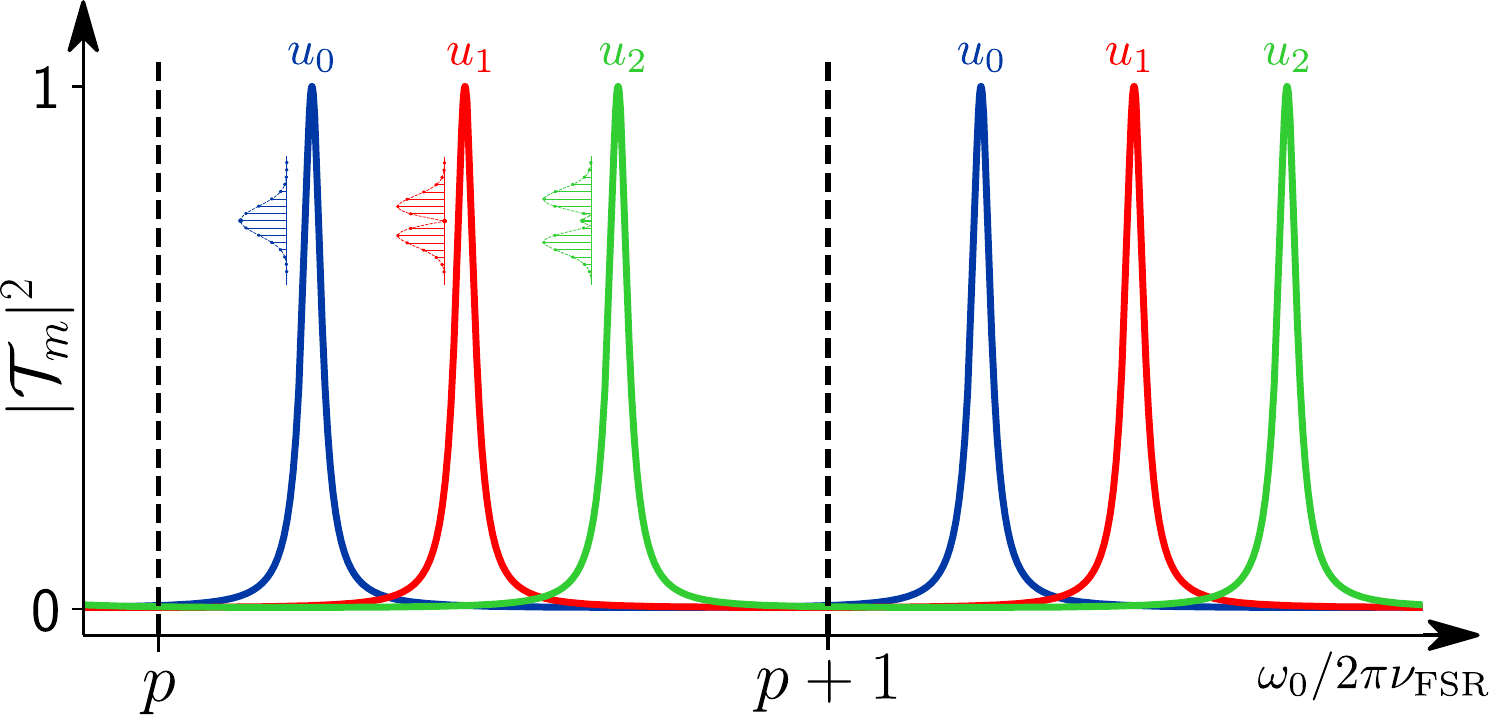}
    \caption{Scheme of the resonances~\eqref{resonances} of a temporal cavity. The integers $p$ and $m$ can be interpreted as the longitudinal and transverse orders of the resonance.}
    \label{fig:resonances}
\end{figure}
In the steady state regime, the mode-dependent input-output relation reads
\begin{equation}
\begin{pmatrix}
\hat{a}_{\mathrm{t},m}^{(k)}
\\
\hat{a}_{\mathrm{r},m}^{(k)}
\end{pmatrix}
=
M
\begin{pmatrix}
\hat{a}_{\mathrm{in}1,m}^{(k)}
\\
\hat{a}_{\mathrm{in}2,m}^{(k)}
\end{pmatrix},
\label{inout}
\end{equation}
where 
\begin{equation}
M
=
\left(
\begin{array}{cc}
\mathcal{T}_m & -\mathrm{e}^{\mathrm{i}\delta_m}\mathcal{R}_m^*
\\
\mathcal{R}_m & \mathrm{e}^{\mathrm{i}\delta_m}\mathcal{T}_m^*
\end{array}
\right),
\label{M}
\end{equation}
with $\mathcal{T}_m=\frac{t_1 t_2 \mathrm{e}^{\mathrm{i}\alpha}}{1-r_1r_2\mathrm{e}^{\mathrm{i}\delta_m}}$ and 
$\mathcal{R}_m=\frac{r_1-r_2 \mathrm{e}^{\mathrm{i}\delta_m}}{1-r_1r_2\mathrm{e}^{\mathrm{i}\delta_m}}$.
In these expressions $\delta_m=k_0 L_{\mathrm{loop}}-\psi_{m}$ is the phase accumulated during one loop and it is the sum of two terms: 
the first is the phase of free space evolution along the full length of the loop $L_{\text{loop}}$ and $\psi_m=(1+m)\psi_{\mathrm{Gouy}}$ 
is the mode-dependent temporal Gouy phase, with $\psi_{\mathrm{Gouy}}=\arctan(D_{\mathrm{tot}}/\sigma_0^2)$, 
accumulated during the propagation through the two dispersive media of total GDD $D_{\mathrm{tot}}=D_1+D_2$. Finally $\alpha$ is the phase accumulated by diffractive propagation from the input coupler to the output coupler.
Since $t_i$ and $r_i$ are chosen real and $t_i^2+r_i^2=1$ (for $i=1,2$), one can easily prove that $M$ is a unitary matrix, 
so that the boson commutation relations for the output field operators are respected.

The temporal cavity effect becomes evident by considering the transmittance coefficient
\begin{align}\label{transmittance}
|\mathcal{T}_m|^2
&=
\frac{\mathcal{T}_{\mathrm{max}}}{1+\left(\frac{2\mathcal{F}}{\pi}\right)^2\sin^2\left(\frac{\omega_0}{2\nu_{\mathrm{FSR}}}-\frac{\psi_{m}}{2}\right)},
\end{align}
where $\mathcal{T}_{\mathrm{max}}=t_1^2 t_2^2/(1-r_1 r_2)^2$, $\mathcal{F}=\pi\sqrt{r_1 r_2}/(1-r_1 r_2)$ and $\nu_{\mathrm{FSR}}=c/L_{\text{loop}}$ 
are the maximal transmittance, the finesse and the free spectral range (FSR) of the temporal cavity, respectively. Mode-dependent resonances are found at
\begin{equation}
    \omega_{(p,m)}=(2\pi p+\psi_m)\nu_{\mathrm{FSR}},
    \label{resonances}
\end{equation} 
with $p\in\mathbb{Z}$. In close analogy with the spatial case, these resonances are characterized by the two integers $p$ and $m$ that can be interpreted as 
the temporal equivalent of the longitudinal and transverse order of a resonance. 

As an example, let us consider an input multimode frequency comb described by expressions~\eqref{E+} and~\eqref{A train} with carrier $\omega_0$, 
repetition rate $T$, and a given temporal $q$-parameter. Injecting these values in eqs.~\eqref{T},~\eqref{evals} and~\eqref{resonances} 
produces three constraints on the cavity parameters $L_{\text{loop}}$, $D_{\mathrm{f}}$ and $D_{\mathrm{tot}}$. 
As a result, for a given input, there is no free parameter for the cavity. Since $L_{\text{loop}}$ and the cavity free spectral range $\nu_{\mathrm{FSR}}$ are blocked, 
via eq.~\eqref{T}, by the train repetition rate, the remaining parameters can be controlled in order to induce the resonance condition for the $m$-th TM and 
at the same time to match the input $q$-parameter. This is an important difference with respect to the spatial-mode cleaner where there is one free parameter, 
for example, the cavity length, that can be used to tune the cavity on a particular resonance. 
This circumstance does not prevent the selection of specific TMs by tuning the temporal cavity to a particular resonance, 
but it certainly makes the operation more challenging. We finally point out that it is possible to free one additional parameter 
by allowing to control the carrier frequency $\omega_0$, the period $T$, or the temporal $q$-parameter of the input.
\\
\indent
\textbf{Temporal cavity as genuine TM mode filter.}
The mode-dependent resonances of a temporal cavity can be employed for filtering multimode frequency combs. 
Let us consider a configuration where a multimode input frequency comb, as in~\eqref{A train}, 
is sent through the first coupler (port ``in1") while keeping the second input (port ``in2") in vacuum state. 
If the cavity is not suitably tuned, the $m$-th order Hermite-Gauss frequency comb is partially transmitted with coefficient $\mathcal{T}_m$ and 
partially reflected with coefficient $\mathcal{R}_m$. As a result, the corresponding quantum state is mixed with the vacuum entering through ``in2". 
Since $M$ is unitary and $|\mathcal{T}_m|^2+|\mathcal{R}_m|^2=1$, tuning the cavity on the suitable resonance $(p,m)$ allows for complete transmission of the mode in the output $\hat{E}_{\mathrm{t}}$,
because $|\mathcal{T}_m|^2=1$ and $|\mathcal{R}_m|^2=0$. As far as the other resonances do not overlap with the one corresponding to $(p,m)$, 
the modes $m'\neq m$ are reflected, at the first coupler, in the mode $\hat{E}_{\mathrm{r}}$, \textit{i.e.} $|\mathcal{T}_{m'}|^2=0$ and $|\mathcal{R}_{m'}|^2=1$. 
Therefore, this configuration allows to filter, or de-multiplex, the mode $m$ from the others without involving any change of the carrier nor of the modal shape of the TM. 

On the other hand, a different configuration can be used to multiplex frequency combs corresponding to different orders. 
For the sake of simplicity let us consider only two modes and suppose that the mode $m$ is sent through the port ``in1", 
while the mode $m'\neq m$ enters the cavity through the port ``in2". When the cavity is tuned on the $m$-th resonance, 
as in the previous configuration, the mode $m$, which is completely transmitted, and the mode $m'$, which is reflected at the second coupler, 
are now joined in the output mode $\hat{E}_{\mathrm{t}}$. This scheme allows therefore to synthesize multimode states with tailored quantum correlations.

\indent
\textbf{Experimental feasibility.}
%
\begin{figure}[t!]
    \centering
    \includegraphics[width=0.8\columnwidth]{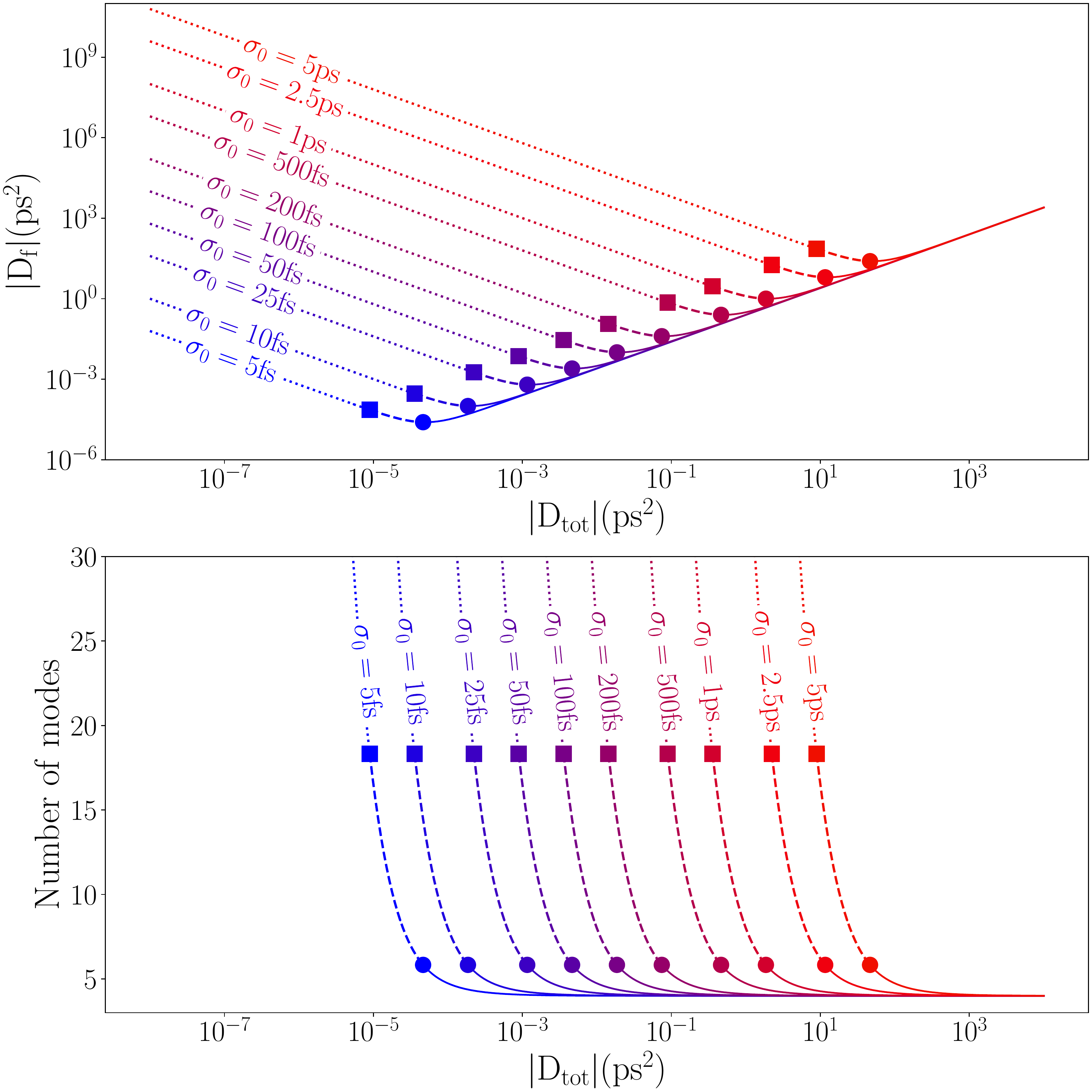}
  \caption{(\textit{Top}) diagram of the relation Eq.~\eqref{eval eq}. (\textit{Bottom}) number of modes fitting one cavity FSR. 
  Full squares and circles correspond to (i) $\mathcal{T}_{\mathrm{max}}=0.77$, $\mathcal{F}=26$, $\mathcal{P}=3\%$ and 
  (ii) $\mathcal{T}_{\mathrm{max}}=0.11$, $\mathcal{F}=8.7$, $\mathcal{P}=40\%$, respectively. 
  At these points, the distance between neighbour resonances is about $1.5$ times their line-width.
  On the left of these points (dotted line for case (i) and dashed line for case (ii)), 
  the distance is smaller and the two resonances are not resolved.}
  \label{fig:constraints}
\end{figure}
The shortest focal lengths experimentally available to date for  efficient electro-optic time lenses are about $10$ ps$^2$ \cite{Torres-Company2011, Karpinski2021}. 
fig.~\ref{fig:constraints}-top depicts the relation Eq.~\eqref{eval eq} constraining the values of $D_{\mathrm{f}}$ and $D_{\mathrm{tot}}$ 
for a given input train of pulses of duration $\sigma_0$. Because of the ``V" shape of these curves, it would always be possible to find a suitable $D_{\mathrm{tot}}$ 
for the experimentally available $D_{\mathrm{f}}$. We note that these values of dispersion are easily achievable \cite{Karpinski2021, Torres-Company2011}. 
However, three factors limit the choice on this parameter: the number $N=2\pi/\psi_{\mathrm{Gouy}}$ of resonances that fit in one cavity FSR, 
the cavity finesse $\mathcal{F}$ and the round-trip losses $\mathcal{P}$. Indeed, two different modes $0\leq m\leq N$ and $m'>N$ cannot be discriminated 
when a resonance of order $(p',m')$ overlaps with the resonance $(p,m)$ (see fig.~\ref{fig:resonances}). 
Therefore $N$ is a figure of merit of a temporal cavity that accounts for the number of modes that can be discriminated. 
In principle, as fig.~\ref{fig:constraints}-bottom shows, $N$ rapidly increases when $D_{\mathrm{tot}}$ decreases. 
In this case, the separation $\psi_{\mathrm{Gouy}}=\arctan(D_{\mathrm{tot}}/\sigma^2)$ between consecutive resonances of 
the same longitudinal order $p$ might become smaller than their line-width $\nu_{\mathrm{FSR}}/\mathcal{F}$. 
Therefore one should seek high values of finesse. On the other side, competing behavior appears when the unavoidable round-trip losses are kept into account. 
These can be modeled as a non-zero reflectance $\mathcal{P}$ of an additional beam splitter, as depicted in Fig.~\eqref{fig:time cav}.
The finesse and the maximal transmittance are now $\mathcal{F}=\pi\sqrt{\rho\, r_1 r_2}/(1-\rho\,r_1 r_2)$ and 
$\mathcal{T}_{\mathrm{max}}=(t_1 t_2)^2/(1-\rho\, r_1 r_2)^2$ respectively, with $\rho^2=1-\mathcal{P}$.
The higher the $\mathcal{F}$, the higher the photon lifetime in the cavity and the sensitivity of $\mathcal{T}_{\mathrm{max}}$ to losses (see Appendix~\ref{Tmax vs finesse} for further details). 
This can rapidly become much smaller than 1 with $\mathcal{P}$ increasing. Therefore a compromise should be sought between $N$ and $\mathcal{F}$ for a given value of losses. 
In figs.~\ref{fig:constraints}, we marked with full squares those points where resonances are resolved with a relative distance of $1.5$ times their line-width for $\mathcal{F}=26$ and $\mathcal{P}=3\%$. 
The corresponding transmittance $\mathcal{T}_{\mathrm{max}}=0.77$ is high enough to guarantee a good efficiency in \mbox{sorting} multimode quantum frequency combs.
On the left of these squares, the dotted parts represent the regions where resonances are not sufficiently resolved. On the contrary,
resonances are well separated on the right side (dashed and solid line regions). 
Here, however, we notice that the value of $N$ decreases and approaches the limit value of 4 (see fig.~\ref{fig:constraints}-bottom).

In the specific case of an electro-optic time lens,
an additional consideration about the time aperture needs to be included: the duration of the pulse at the time lens needs to be shorter than the span of the parabolic phase modulation. 
Assuming the standard cosinusoidal electro-optic time lens, where $D_f = 4\pi^2  \delta f_{\mathrm{RF}}^2$ ($\delta$ -- modulation amplitude, $f_{\mathrm{RF}}^2$ -- modulation frequency) 
with an experimentally realistic value of $\delta = 11~\mathrm{rad}$, and the temporal aperture 
of $0.4/f_{\mathrm{RF}}$ \cite{Torres-Company2011,Karpinski2021, Kapoor2022}, 
we verified that this condition is satisfied for the maximum pulse durations allowed by eq.~\ref{evals}. 
We considered the condition as met when double FWHM of the pulse was smaller than the aperture. 
We further note that 40\% loss is feasible to realize given the losses in the state of the art of electro-optic and dispersive elements. 
For this level of losses, the parameters region for which resonances are resolved is on the right side of the full circles in figs.~\ref{fig:constraints}.
For these points resonances are separated about $1.5$ times their line-width and the corresponding transmittance $\mathcal{T}_{\mathrm{max}}=0.11$ is high enough to be detected in an experimental test. 
The development of integrated thin-film lithium niobate devices \cite{Zhu2022} is expected to further increase the experimental robustness of the scheme. 

\indent
\textbf{Conclusions}
We propose a method to realize a genuine temporal mode filter for frequency combs, which transmits a single temporal mode while blocking the other temporal modes. 
The filter operation is based on the temporal mode dependence of the temporal Gouy phase combined with a cavity build-up effect. 
We show that the implementation of the filter is experimentally feasible using currently available electro-optic time lenses. 
The filter operation does not rely on complex nonlinear interactions or phase matching, is frequency independent and does not change the temporal mode. 
It will facilitate new applications in multidimensional quantum information processing and time-frequency metrology by enabling multiplexing and demultiplexing of temporal modes, 
including temporal mode-dependent detection. In particular it will enable the synthesis of multimode quantum frequency combs as resources for quantum networks~\cite{Cai2017} 
as well as for quantum metrology~\cite{Lamine2008}.

\begin{acknowledgments}
We are grateful to Nicolas Treps and Valentina Parigi for the frutiful discussions.
This work was partially supported by the network QuantERA of the European Union’s Horizon 2020 research and innovation program under project 
“Quantum information and communication with high-dimensional encoding” (QuICHE), by the Agence Nationale de la Recherche through LABEX CEMPI (ANR-11-LABX-0007) and 
I-SITE ULNE (ANR-16-IDEX-0004) and by the National Science Centre of Poland (Project No. 2019/32/Z/ST2/00018), 
by Lille University through the program ``Internationalisation de la recherche 2022 -- collaboration bilat\`{e}rales".
\end{acknowledgments}

\section*{Appendices}
\appendix
\section{Gaussian temporal imaging}\label{gauss imag}
In a quantum description of ultra-fast optical pulses, the positive-frequency part of the electric field operator can be written as~\cite{Fabre2020}
\begin{align}
\hat{E}^{(+)}(t,z)&=\mathcal{E}_0
\mathrm{e}^{\mathrm{i}\left(k_0 z-\omega_0 t\right)}
\hat{A}(t,z),
\label{E+}
\\
\hat{A}(t,z)&=
\sum_{m\in\mathbb{N}}\,\hat{a}_m u_m(t,z),
\label{A}
\end{align}
where $\omega_0$ is the carrier, $k_0=k(\omega_0)$, $\mathcal{E}_0$ is the single photon amplitude of the mode $m$ and $\hat{a}_m$ 
are annihilation operators satisfying boson commutation relations $[\hat{a}_m,\hat{a}^\dag_n]=\delta_{m,n}$.
Here we assume a pulse propagating along the $z$ and having a trivial profile in the transverse plane $(x,y)$, i.e. a plane wave. 
The slowly-varying envelope operator $\hat{A}(t,z)$ is decomposed over a generic orthonormal basis of modal functions $\{u_m(t,z)\}_{m\in\mathbb{N}}$. 
It is convenient working in the travelling-wave frame of reference $(\tau,z)$ that is propagating with the wave at pulse group velocity where the 
\textit{delayed time} $\tau$, defined as $\tau=t-\beta_{1} z$, with $\beta_{1}=(\mathrm{d} k/\mathrm{d}\omega)_{\omega_0}$ the inverse of group velocity.
In the quasi-monochromatic approximation, their propagation through a dispersive medium is described by the equation
\begin{align}
\frac{\partial}{\partial z}u_m(\tau,z)
&=
-\frac{\mathrm{i}}{2}\beta_2\frac{\partial^2}{\partial\tau^2}u_m(\tau,z)
\label{disp evo}
\end{align}
where $\beta_{2}=(\mathrm{d}^2k/\mathrm{d}\omega^2)_{\omega_0}$ is the Group Velocity Dispersion (GVD).
Since eq.~\eqref{disp evo} is the analogous of the paraxial Helmholtz equation for diffracting beams, the orthonormal basis of temporal Hermite-Gauss (HG) functions is a possible solution.
Up to a normalization factor, these solutions read, for $m\in\mathbb{N}$, as 
\begin{align}
u_m(\tau,z)\propto \frac{\sigma_0}{\sigma(z)}
H_m\left(\tau/\sigma(z)\right)
\mathrm{e}^{-\mathrm{i}\tau^2/(2q(z))}
\mathrm{e}^{-\mathrm{i}(1+m)\psi(z)}
\label{GH}
\end{align}
with $H_m(\tau/\sigma(z))$ the Hermite functions, $1/q(z)=C(z)-\mathrm{i}/\sigma^2(z)$ the inverse of the temporal complex beam parameter $q$, $\sigma(z)=\sigma_0(1+(\beta_2 z/\sigma_0^2)^2)^{1/2}$ 
the pulse duration at the propagation distance $z$, $C(z)^{-1}=\beta_2 z(1+(\sigma_0^2/(\beta_2 z)^2))^{1/2}$ the inverse of the chirp acquired by the pulse at $z$ because of dispersion, 
$\psi(z)=\arctan(\beta_2 z/\sigma_0^2)$ the temporal Gouy phase shift, and $\sigma_0$ the minimal pulse duration reached when its chirp is null. 
The $q$ parameter, as is the case for its spatial analogue, contains all the information about the Hermite-Gaussian pulse.

Because of the space-time duality, the evolution of Hemite-Gauss temporal modes (HGTMs) can be solved by using the temporal version of the ABCD formalism~\cite{Nakazawa1998}. 
Therefore, for $z_2>z_1$ the $q$-parameter is given by
\begin{align}
q(z_2)=\frac{A q(z_1) + B}{C q(z_1) + D},
\end{align}
where the coefficients $A$, $B$, $C$ and $D$ are the matrix elements of the equivalent ray transfer matrix of all the optical elements through which the pulse has travelled from $z_1$ 
to $z_2$ and they satisfy the relation $AD-BC=1$. In this formalism, the ABCD matrix $M_{\text{disp}}$ for a dispersive propagation through a medium of length $d$ and GVD $\beta_2$ is given by
\begin{align}
M_{\text{disp}}&=
\begin{pmatrix}
1 & D
\\
0 & 1
\end{pmatrix},
\end{align}
where $D=\beta_2 d$ is the group delay dispersion (GDD). In a similar way, by invoking the space-time duality, we obtain the ABCD matrix $M_{\text{TL}}$ of a time lens of focal-GDD $D_{\mathrm{f}}$
\begin{align}
M_{\text{TL}}&=
\begin{pmatrix}
1 & 0
\\
-1/D_{\mathrm{f}} & 1
\end{pmatrix}.
\end{align}
Equipped with these few elements we develop the theory of a temporal cavity in the main text.

\section{Sensitivity of the maximal transmittance to finesse.}\label{Tmax vs finesse}
\begin{figure}[t!]
    \centering
    \includegraphics[width=\columnwidth]{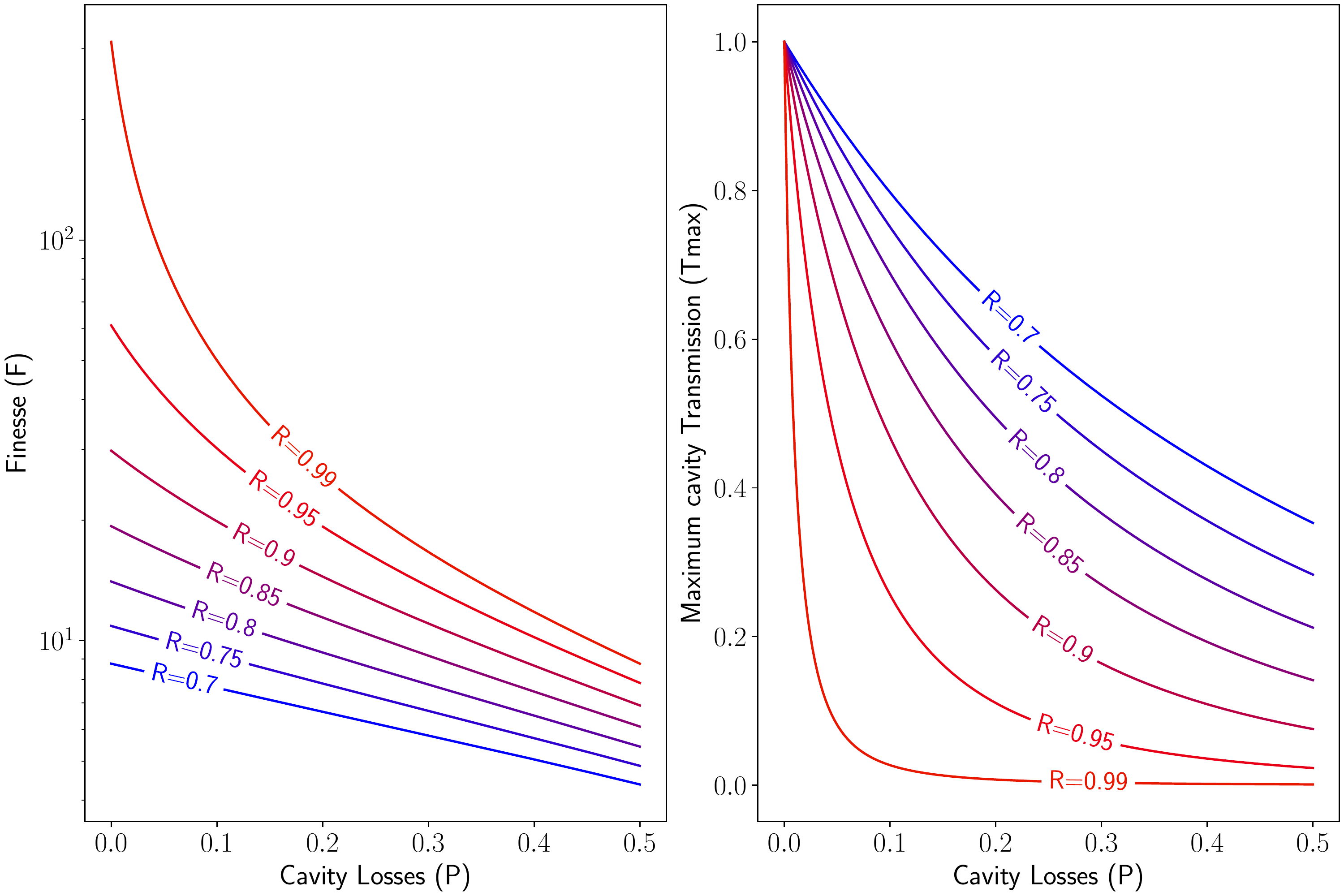}
  \caption{(\textit{Left}) Evolution of the maximal transmittance $\mathcal{T}_{\mathrm{max}}$ (eq.~\ref{eq:Tmax}) and (\textit{right}) the finesse $\mathcal{F}$ (eq.~\ref{eq:F}) 
  for different values of the reflectance of the beamsplitters $R$.}
  \label{fig:tmax_finesse}
\end{figure}
When the unavoidable intracavity losses (say $\mathcal{P}$) are kept into account, the finesse $\mathcal{F}$ and 
the maximal transmittance $\mathcal{T}_{\mathrm{max}}$ of the temporal cavity can be quantified as follows:
\begin{align}
\mathcal{T}_{\mathrm{max}} &= \frac{(1-\mathcal{R})^2}{(1-\mathcal{R}\sqrt{1-\mathcal{P}})^2},
\label{eq:Tmax}
\\
\mathcal{F} &= \frac{\pi\sqrt{\mathcal{R}\sqrt{1-\mathcal{P}}}}{1-\mathcal{R}\sqrt{1-\mathcal{P}}}.
\label{eq:F}
\end{align}
Here $\mathcal{R}=r_1^2 =r_2^2$ denotes the reflectivity of the input and output couplers, which can be adjusted to achieve the desired level of finesse and maximal transmittance.
As noted in the main text, a higher value of the finesse is desirable for achieving a better separation between the mode-dependent resonances.
However, a high finesse leads to a longer photon lifetime in the cavity that induces an increased sensitivity of the maximal transmittance to the cavity losses. 
This trade-off is illustrated in fig.~\ref{fig:tmax_finesse}. In the left panel, for a given value of losses $\mathcal{P}$,
the finesse can be increased (reduced) by increasing (reducing) the reflectivity $\mathcal{R}$ of the input-output couplers.
On the contrary, on the right, an increase of the reflectivity $\mathcal{R}$ leads to a decrease of $\mathcal{T}_{\mathrm{max}}$.

As losses increase, the value of $\mathcal{T}_{\mathrm{max}}$ decreases accordingly, and this effect is more pronounced for higher values of $\mathcal{R}$, 
as seen from the steeper drop in $\mathcal{T}_{\mathrm{max}}$ observed at these reflectivities. 

As a result, the choice of reflectivity for the beamsplitters must balance out the desire for a high finesse and therefore longer photon lifetime with the need for 
a high maximal transmittance and therefore greater efficiency. 
This results in a trade-off between the two metrics that must be carefully considered when designing temporal mode filters.

For classical applications, where pulse amplification is possible, the value of $\mathcal{T}_{\mathrm{max}}$ is generally not critical. 
In such cases, the choice of $\mathcal{R}$ should prioritize a high degree of finesse, as it corresponds to better separation between cavity modes. 
However, in quantum applications, the sensitivity of Tmax to losses becomes a much more important consideration, since it determines the efficiency of the filtered mode. 
Thus, the choice of $\mathcal{R}$ for such applications need to ensure maximizing the efficiency of the cavity while maintaining a sufficient level of finesse. 
That will depend on the overall losses in the cavity.

\bibliography{temporal_cavity}


\end{document}